\documentclass[sigconf]{acmart}
\renewcommand\footnotetextcopyrightpermission[1]{} 
\usepackage{enumitem}
\AtBeginDocument{%
  \providecommand\BibTeX{{%
    \normalfont B\kern-0.5em{\scshape i\kern-0.25em b}\kern-0.8em\TeX}}}

\begin{document}

\title[Exploring The Role of Local and Global Explanations in Recommender Systems]{Exploring The Role of Local and Global Explanations in Recommender Systems}

\author{Marissa Radensky}
\authornote{Work conducted during internship at the Allen Institute for Artificial Intelligence and Ph.D. at the University of Washington.}
\email{radensky@cs.washington.edu}
\affiliation{%
  \institution{University of Washington}
  \country{}
}
\author{Doug Downey}
\email{dougd@allenai.org}
\affiliation{%
  \institution{Allen Institute for Artificial Intelligence \& Northwestern University}
  \country{}
}
\author{Kyle Lo}
\email{kylel@allenai.org}
\affiliation{%
  \institution{Allen Institute for Artificial Intelligence}
  \country{}
}
\author{Zoran Popović}
\email{zoran@cs.washington.edu}
\affiliation{%
  \institution{University of Washington}
  \country{}
}
\author{Daniel S. Weld}
\email{weld@cs.washington.edu}
\affiliation{%
 \institution{University of Washington \& Allen Institute for Artificial Intelligence}
 \country{}
}


\begin{abstract}
Explanations are well-known to improve recommender systems' transparency. These explanations may be local, explaining an individual recommendation, or global, explaining the recommender model in general. Despite their widespread use, there has been little investigation into the relative benefits of these two approaches. Do they provide the same benefits to users, or do they serve different purposes? We conducted a 30-participant exploratory study and a 30-participant controlled user study with a research-paper recommender system to analyze how providing participants local, global, or both explanations influences user understanding of system behavior. Our results provide evidence suggesting that both explanations are more helpful than either alone for explaining how to {\em improve} recommendations, yet both appeared less helpful than global alone for efficiency in {\em identifying} false positives and negatives. However, we note that the two explanation approaches may be better compared in the context of a higher-stakes or more opaque domain.
\end{abstract}



\maketitle
\pagestyle{plain} 

\section{Introduction}
Recommender systems are used daily by millions of people, and explanations that clarify a recommender's behavior are well-known to improve users' perceptions of the recommender's usefulness \cite{Bakalov2013AnAT,Bostandjiev2012TasteWeightsAV,Bostandjiev2013LinkedVisES,Tsai2019ExplainingRI,Knijnenburg2012InspectabilityAC,Felfernig2006AnES,Vig2012TheTG,Dominguez2019TheEO,tsai2020effects}, controllability \cite{Bakalov2013AnAT,Gretarsson2010SmallWorldsVS,Knijnenburg2012InspectabilityAC,Lee2020ExplanationBasedTO}, trustworthiness \cite{Bruns2015WhatSI,Ribeiro2016WhySI,Ahn2007OpenUP,Li2019WhatDS,Lee2020ExplanationBasedTO,Felfernig2006AnES}, and transparency \cite{Bakalov2013AnAT,Chang2019SearchLensCA,Gretarsson2010SmallWorldsVS,Li2019WhatDS,Knijnenburg2012InspectabilityAC,Tsai2017ProvidingCA,Kulesza2012TellMM,Lee2020ExplanationBasedTO}. Some recommenders provide users with {\em local} explanations describing why a specific item is recommended \cite{Cramer2008TheEO,Lee2020ExplanationBasedTO}. Others give users a {\em global} explanation describing how recommendations are selected by the system overall \cite{ODonovan2008PeerChooserVI,Kangasrsi2015ImprovingCA}. Still others show {\em both} explanations, which can be presented separately \cite{Parra2015UsercontrollablePA,Ahn2007OpenUP,Knijnenburg2011EachTH,Tsai2019ExplainingRI,Bakalov2013AnAT,Jin2018EffectsOP,Millecamp2019ToEO} or in a unified manner \cite{Gretarsson2010SmallWorldsVS,Bostandjiev2012TasteWeightsAV,Bostandjiev2013LinkedVisES,Bruns2015WhatSI,Schaffer2015HypotheticalRA,Devendorf2012TopicLensA}. 
 
Despite the widespread use of local and global explanations in recommender systems, to the best of our knowledge there has been no investigation into how each type of explanation influences the transparency of a recommender system. Recommenders often require feedback in order to provide high quality recommendations. Do the two explanation types play complementary roles in helping users understand how the system may improve recommendations? Are local explanations used differently if global explanations are also present, or vice versa? Is one explanation type better for detecting false positive or false negative recommendations? We examine these questions and more using the recommender Semantic Sanity, which allows users to create recommendation feeds of computer-science research papers. 

In summary, we make the following contributions:
\begin{itemize}
  \item A formative study regarding how to present local and global explanations in a research-paper recommender.
  \item An exploratory study and controlled user study, each with 30 computer-science researchers, using Semantic Sanity to investigate several hypotheses surrounding three conditions: local, global, and local-plus-global explanations.
  \item Evidence suggesting that 1) both explanations help users explain how to improve recommendations better than either alone, but 2) both is less helpful than global alone for efficiency in identifying false positives and negatives. Also, 3) users prefer less diverse local explanations when a global explanation is also available.
\end{itemize}

\section{Related Work}

\subsection{Local and Global Explanations in Machine Learning}
In machine learning broadly, global explanations explain how a model behaves generally, while local explanations explain a single model output, as first distinguished by Ribeiro et al. \cite{Ribeiro2016WhySI}. With respect to model transparency, local and global explanations have been studied from several perspectives. Some works find that local explanations have advantages over global explanations. Ribeiro et al. \cite{Ribeiro2016WhySI} established that local explanations more easily achieve model faithfulness. Similarly, Guidotti et al. \cite{Guidotti2018LocalRE} found that local explanations are more accurate and less complex than global explanations in simulating a black-box model's decisions. Other studies discuss benefits from both local and global explanations. For an image classification task, Mishra et al. \cite{mishra2021crowdsourcing} observed that local and global explanations both aid users in estimating model confidence and gauging their own confidence in the model output. Huber et al. \cite{huber2020local} found that participants shown both local and global explanations instead of either alone performed best in evaluating reinforcement-learning agents. For a task predicting risk of recidivism, one study demonstrated that local explanations are more helpful in discerning algorithmic fairness on a case-by-case basis, yet global explanations are perceived as more useful for understanding the model \cite{Dodge2019ExplainingMA}. Another study showed that data scientists found both local and global explanations useful in trying to understand a model. However, novices preferred local explanations, while experts preferred global explanations \cite{Hohman2019GamutAD}. In addition, Kopitar et al. \cite{kopitar2019local} saw evidence that local interpretability provides additional insight over global interpretability in machine learning models for type 2 diabetes mellitus screening. We build on these works to address local and global explanations for transparency of recommender systems in particular. Recommender systems differ from most AI systems in that their output cannot be objectively evaluated as correct or not. Local and global explanations may be used differently when users must \emph{subjectively} decide whether or not a recommendation is good and utilize that information to provide feedback to the recommender.

\subsection{Explanations for Appropriate Trust of AI}
\label{sec:trust}
When someone has appropriate trust in an intelligent system, they recognize when it is correct or not \cite{lee2004trust}. Studies have shown that appropriate trust of AI can be difficult to attain through explanations \cite{van2021evaluating,mishra2021crowdsourcing, chu2020visual,hase2020evaluating,bansal2020does,zhang2020effect,jacobs2021machine}. When the AI and user have similar decision-making performance, two studies found that, when compared to AI confidence, explanations do not improve team performance \cite{bansal2020does} or trust calibration \cite{zhang2020effect} respectively. However, in a study surrounding a question-answering task, explanations did help users develop more appropriate trust in comparison to AI confidence \cite{gonzalez2020human}. The authors note this difference may be caused by the fact that, unlike other tasks, this one's explanations provide users with previously unseen information rather than just weighting of already seen evidence. Another study found that the timing and presentation of explanations can encourage more or less appropriate trust, but users prefer settings inducing less appropriate trust \cite{buccinca2021trust}. With regards to local and global explanations, Huber et al. \cite{huber2020local} saw that both may help establish appropriate trust in reinforcement-learning agents. Meanwhile, Mishra et al. \cite{mishra2021crowdsourcing} found that, in an image classification task, global explanations were slightly less helpful for estimating model confidence in true positives compared to false positives. Here, we explore how local and global explanations may influence appropriate trust of a recommender by investigating if they help users to identify false positive and false negative recommendations.

\subsection{Dimensions of AI Explanations}
AI explanations have been designed and studied along several dimensions in addition to that of local and global explanations. For one, they may be generated using model-agnostic \cite{Ribeiro2016WhySI,lundberg2017unified} or model-specific \cite{lundberg2017unified,martens2008rule,landecker2013interpreting} methods. They may be factual, explaining why a certain model outcome occurred, or contrastive, explaining why another outcome did not occur; they may also be counterfactual, explaining how another outcome could have occurred instead \cite{guidotti2019factual,stepin2021survey,hendricks2018generating}. Furthermore, they have been investigated with regards to diverse user attributes such as their domain expertise \cite{schaffer2019can,mohseni2018survey,ribera2019can}, machine-learning expertise \cite{Hohman2019GamutAD,szymanski2021visual,mohseni2018survey,ribera2019can}, stakeholder group (e.g., developers, end users) \cite{preece2018stakeholders}, cognitive skills \cite{Millecamp2019ToEO}, and personality traits (e.g., openness, neuroticism) \cite{Kouki2019PersonalizedEF,Millecamp2019ToEO}. AI explanations have additionally been compared in terms of various modalities such as visualization \cite{gonzalez2020human,szymanski2021visual}, text \cite{szymanski2021visual}, and audio \cite{gonzalez2020human}. Other studies have varied the length \cite{gonzalez2020human} and number \cite{Kouki2019PersonalizedEF} of explanations to observe the impact on cognitive load, as well as the toggle-ability and timing of explanations to reduce user biasing \cite{buccinca2021trust}. Two explanation dimensions that have been studied specifically in the context of recommender systems are style (e.g., social-based, content-based) \cite{Kouki2019PersonalizedEF,Friedrich2011ATF} and actionability \cite{Vig2012TheTG,Kangasrsi2015ImprovingCA,Knijnenburg2011EachTH,Lee2020ExplanationBasedTO,Nart2013PersonalizedAT}. In this paper, the local and global explanations are content-based and actionable.

\begin{figure}[tb]
  \centering
  \includegraphics[width=8.5cm]{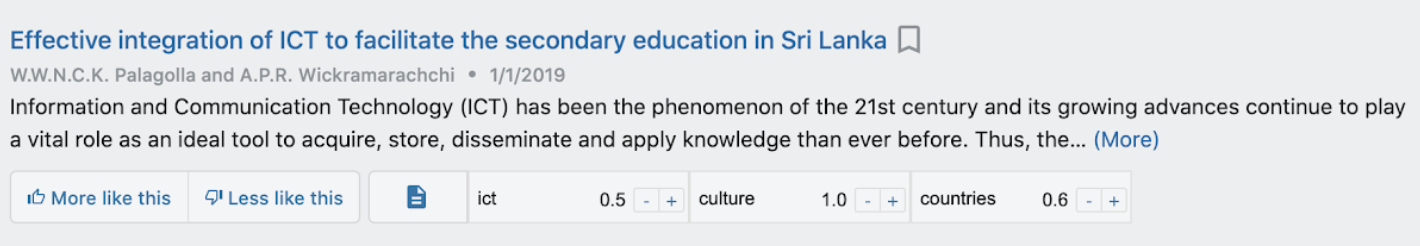}
  \vspace*{-.1in}
  \caption{A paper recommendation in the local condition of Study 2, with the local explanation open at the bottom.}
  \label{fig:tutLocal}
  \Description{A paper recommendation is shown with the title "Effective integration of ICT to facilitate the secondary education in Sri Lanka." The title is light blue to indicate it is a link. There is a bookmark button to the right of the title to save the recommendation. The authors of the paper and the publication date are shown directly underneath, followed by the first couple lines of the abstract. More of the abstract is available if the user clicks the button that says "(More)" at the end of those two lines. Underneath all of that, there are two buttons saying "More like this" with a thumbs up icon before the words and "Less like this" with a thumbs down icon before the words. To the right of these buttons, there is a button with a document icon that is darkened to indicate it is pressed. Directly connected to this button to the right are three term boxes. The term boxes contain a term followed by a weight and two buttons showing a "-" and a "+" respectively. The three terms and weights are "ict" - 0.5, "culture" - 1.0, and "countries" - 0.6.}
\end{figure}

\begin{figure*}[ht]
  \centering
  \includegraphics[width=8.5cm]{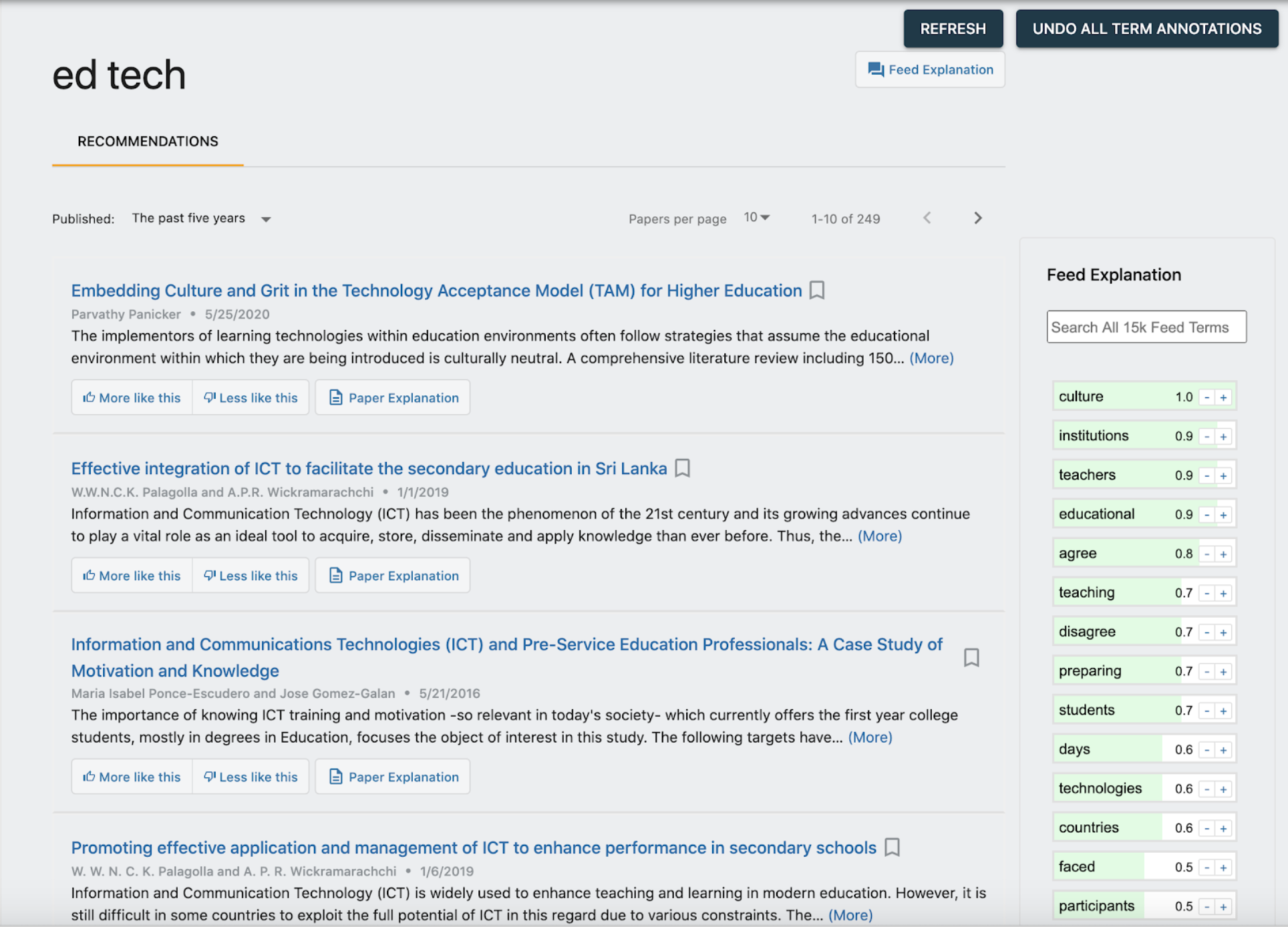}\hfill
  \includegraphics[width=8.5cm]{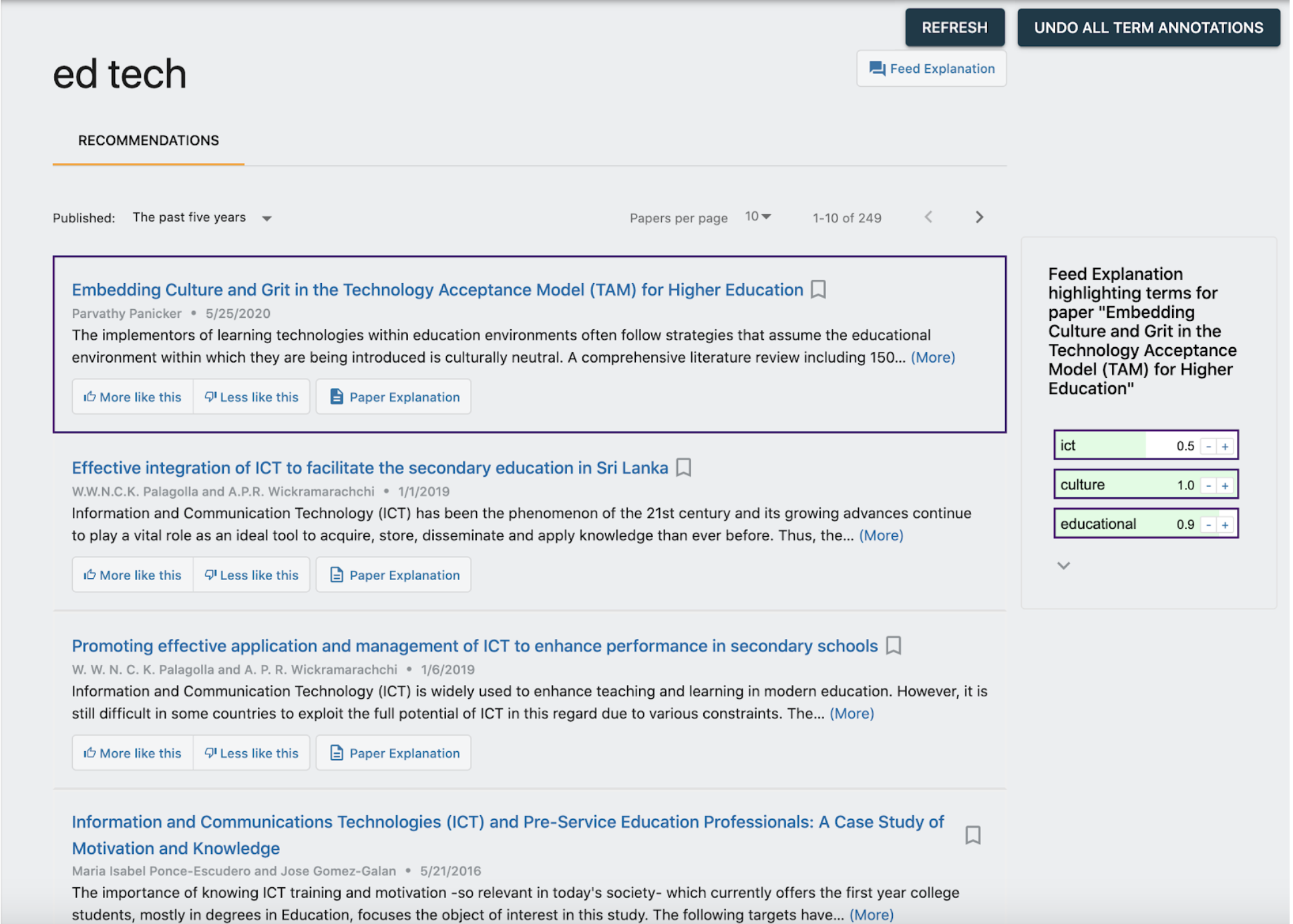}
  \vspace*{-.1in}
  \caption{User interface for the local-plus-global condition in Study 2. Left: default layout. Right: layout when a local explanation is open. Irrespective of condition, the following features are present: "(More)" button under each paper to see its full abstract, "More like this"/"Less like this" buttons under each paper to provide feedback, a bookmark button next to each paper to save it, a "Refresh" button to apply user feedback, an "Undo Term Annotations Applied By Refresh" button shown directly after refreshing to undo all term annotations applied in the last refresh, and an "Undo All Term Annotations" button to return all terms to their original ratings.}
  \label{fig:tutBoth}
  \Description{In both images, the feed is titled "ed tech" at the top. Underneath that, there is one tab for recommendations with an orange line underneath the word "Recommendations," indicating that tab is open. Underneath the "Recommendations" tab header, there is a "Published" filter which is set for "The past five years." To the right of this filter, there is a drop down menu for "Papers per page" that is set to 10. To the right of that, "1-10 of 249" is written to indicate the top 10 recommendations out of the top 249 recommendations are being shown on this page. To the right of that, there are left and right arrows to page through the 249 recommendations. The left arrow is thus grayed out. In the top right corner of the images, there is a button that says "Feed Explanation" with a chat icon before it. The chat icon is darkened to indicate the button is pressed. Directly above that button, a "REFRESH" button is shown. To the right of that button, An "UNDO ALL TERM ANNOTATIONS" button is shown. Four recommendations are shown: "Embedding Culture and Grit in Technology Acceptance Model (TAM) for Higher Education," "Effective Integration of ICT to facilitate the secondary education in Sri Lanka," "Information and Communications Technologies (ICT) and Pre-Service Education Professionals: A Case Study of Motivation and Knowledge," and "Promoting effective application and management of ICT to enhance performance in secondary schools." Each recommendation presents the paper title, authors and publication date underneath, and a bit of the abstract underneath that, which can be seen in full by clicking "(More)." There is a bookmark button to the right of each title to save the paper. The paper titles are light blue indicating that they are links. At the bottom of each recommendation, there are three buttons that say: "More like this" with a thumbs up icon before the words, "Less like this" with a thumbs down icon before the words, and "Paper Explanation" with a document icon before the words. To the right of the user interface, the Feed Explanation bar is shown. It is titled "Feed Explanation." In the left image specifically, under the title of the Feed Explanation bar, there is a search bar that reads "Search All 15k Feed Terms." Underneath the search bar in the Feed Explanation, 14 term boxes are visible with weights to their right, and a green bar partially fills each term box to represent its weight between 0.0 and 1.0. The terms and weights are, in descending order: "culture" - 1.0, "institutions" - 0.9, "teachers" - 0.9, "educational" - 0.9, "agree" - 0.8, "teaching" - 0.7, "disagree" - 0.7, "preparing" - 0.7, "students" - 0.7, "days" - 0.6, "technologies" - 0.6, "countries" - 0.6, "faced" - 0.5, and "participants" - 0.5. To the right of each term and weight, a "-" button followed by a "+" button are shown. In the right image specifically, the first recommendation is highlighted with a dark purple border. The Feed Explanation bar's title now says "Feed Explanation highlighting terms for paper "Embedding Culture and Grit in Technology Acceptance Model (TAM) for Higher Education." Three term boxes are shown beneath this title: "ict," culture," and "educational." The terms have weights of 0.5, 1.0, and 0.9 respectively. All three terms are highlighted with a dark purple border. A green bar partially fills each term box to represent its weight between 0.0 and 1.0, and to the right of each term and weight, a "-" button followed by a "+" button are shown. Underneath the terms, a carrot button faces downward, indicating the user can click it to expand the sidebar to show the highlighted terms in context of all the feed's top terms.}
\end{figure*}

\begin{table*}[ht]
  \caption{Metrics considered in Study 2, with corresponding hypotheses defined in Section \ref{sec:hypotheses}. The questions are 7-point Likert-type questions. The two log file metrics (LFM) are from a click log file.}
  \centering
  \label{tab:metrics}
  \begin{tabular}{p{0.045\linewidth}p{0.17\linewidth}p{0.72\linewidth}}
    \toprule
    \textbf{Hypo.} & \textbf{Metric ID} & \textbf{Metric}\\
    \midrule
    -& Q0: feed success &"The recommendation feed helps me find relevant papers."\\
    H1& Q1: past actions &"The explanation(s) help me to understand why the system returned the papers it did."\\
    H2& Q2: future actions &"The explanation(s) help me to anticipate what kinds of papers the system will return in the future."\\
    H3& Q3: understand me &"The explanation(s) help me to know when the system doesn’t understand my interests."\\
    H4& Q4: change behavior &"When the feed is not completely relevant, I can explain how I would like the system to behave to be more relevant."\\
    H5& Q5: false pos paper &"The explanation(s) help me to determine whether a \textbf{paper} is relevant or irrelevant."\\
    & Q6: false pos term &"The explanation(s) help me to understand which \textbf{term} might cause an irrelevant paper to appear in my feed."\\
    &LFM1&\% of annotated terms that are annotated negatively\\
    H6& Q7: false negative &"The explanation(s) help me to understand how likely the feed is to \textbf{miss papers} that I’d consider relevant."\\
    H7& Q8: local diversity &"I would like the Paper Explanations to cover a less diverse set of terms, focusing more on the highest-rated terms."\\
    H8&LFM2&\# of annotated terms\\
  \bottomrule
\end{tabular}
\end{table*}

\section{Study 1: Formative Study for System Design}

We first ran a formative study presenting design mockups for the recommender Semantic Sanity to six computer-science researchers in order to determine how best to present local, global, and local-plus-global explanations. These explanations are terms (unigrams and bigrams) from papers, a form of the common content-based explanation ~\cite{Kouki2019PersonalizedEF,Friedrich2011ATF, Bakalov2013AnAT, Bostandjiev2012TasteWeightsAV,Jin2018EffectsOP, Ahn2007OpenUP}. The global terms are those with the most positive weights in the linear model for selecting paper recommendations. The local terms are those with the most positive product of model weight and TF-IDF value for the term's associated paper, and we use LIMEADE's approach \cite{Lee2020ExplanationBasedTO} for introducing some randomness to diversify the local terms. We found a majority of participants preferred that local and global explanations be toggle-able and that they be presented in a unified manner when both available. Most participants also desired that they be actionable, meaning the user may directly manipulate the explanation widget to provide feedback to the recommender system \cite{Lee2020ExplanationBasedTO}. Furthermore, participants easily understood that when local explanations had varying numbers of terms, only the most significant terms were shown, so we allowed variable-length local explanations. Within the constraint of two to four terms total, the system added terms to the local explanation until the term weights hit a plateau, meaning the explanation had the most salient terms.

Figure~\ref{fig:tutBoth} shows the resulting interface for the local-plus-global condition. In all conditions, users can like or dislike papers and give feedback on terms considered by the model. In the {\bf local-plus-global} condition, the "Feed Explanation" button at the top allows users to close or reopen a sidebar containing the global explanation. The sidebar presents the top 80 terms most related to the feed and allows users to search all 15,000 terms. Users can adjust terms' importance to the feed using the plus and minus buttons. The user may adjust terms' ratings between 0.0 and 1.0; one click adds or subtracts 0.1 to a term's rating. Additionally, users can click the "Paper Explanation" button under each paper to display a local explanation. This surfaces two to four paper-relevant terms at the top of the sidebar, and users can click the carrot underneath them to put the terms in context of the feed's other top terms. The {\bf global} condition looks similar but does not include the "Paper Explanation" buttons. In the {\bf local} condition, users can click the "Paper Explanation" button under each paper to reveal or hide two to four terms explaining why the paper was recommended (Figure \ref{fig:tutLocal}). They can also open all local explanations with a "View All Paper Explanations" button.

\section{Study 2: Exploratory Study}

\subsection{Study Design}
\subsubsection{Hypotheses}
\label{sec:hypotheses}
The objective of Study 2 was to explore how people use local and global explanations in a research-paper recommender system. The first six hypotheses concern transparency and are inspired by target purposes of AI explanations enumerated in previous work \cite{Hoffman2018MetricsFE,gunning2017explainable}. These hypotheses state that there is at least one paired difference among the local, global, and local-plus-global conditions with regards to how helpful they are for... 
\textbf{H1}: understanding the recommender’s past actions, \textbf{H2}: understanding the recommender's future actions, \textbf{H3}: knowing how well the system understands the user, \textbf{H4}: understanding how the system can improve, \textbf{H5}: identifying false positives, and \textbf{H6}: identifying false negatives.
The final two hypotheses address how users' interactions with the explanations are affected by the explanation types provided. 
\textbf{H7}: There is a difference between local and local-plus-global with regards to how diverse users want the local explanation terms to be, and \textbf{H8}: there is at least one paired difference among local, global, and local-plus-global with regards to how much feedback users provide on their explanations.
The hypotheses' related 7-point Likert-type questions and log file metrics are outlined in Table \ref{tab:metrics}. 

\subsubsection{Participants and Treatments}
Thirty researchers who read at least one computer-science research paper each month interacted with the recommender system in a half-hour to one-hour session and were compensated with \$25 Amazon gift cards.
Fifteen participants went through both the global and local conditions in randomized order, and the other 15 interacted only with the local-plus-global condition. We did not include a baseline condition (no explanation) because the importance of explanations to recommender transparency is well-established \cite{Bakalov2013AnAT,Chang2019SearchLensCA,Gretarsson2010SmallWorldsVS,Li2019WhatDS,Knijnenburg2012InspectabilityAC,Tsai2017ProvidingCA,Kulesza2012TellMM,Lee2020ExplanationBasedTO}. When signing up for the study, each participant provided two topics of interest, which would act as their feed topics. 

\subsubsection{Procedure}
We first presented participants with a condition-specific slide tutorial. We then instructed participants to navigate to a specified link in order to access the recommender system. Clicks during the interaction with the system were recorded in a log file. Next, participants started their recommendation feed about their preset feed topic by selecting 4 seed papers, found using keyword search. Once they narrowed down their seed papers, they named and generated the feed. The participants' objective was to make the recommendation feed as relevant to them as possible. They had 15 minutes to do so, but if they felt that the feed was not going to become any more relevant before 15 minutes had passed, they stopped early. We also asked participants to think aloud as they interacted with the system in case there were any helpful insights into their interactions or they needed a reminder of how to use a certain system feature.

At the end of each condition, participants filled out a Google Forms survey without looking at the system. The survey first asked for short answers regarding in what situations, if any, the participant found each type of explanation useful. If the participants had any other thoughts on the explanations, they provided those as well. After, they answered the Likert-type questions discussed in Table \ref{tab:metrics}. Lastly, participants returned to their feed and categorized the final top ten papers as relevant, neutral, or irrelevant. However, since this data depended heavily on factors other than successful feed curation (e.g. the number of papers published on the feed topic), we did not utilize it.

\subsection{Results and Discussion}

\begin{figure*}[ht]
  \centering
  \includegraphics[width=13cm]{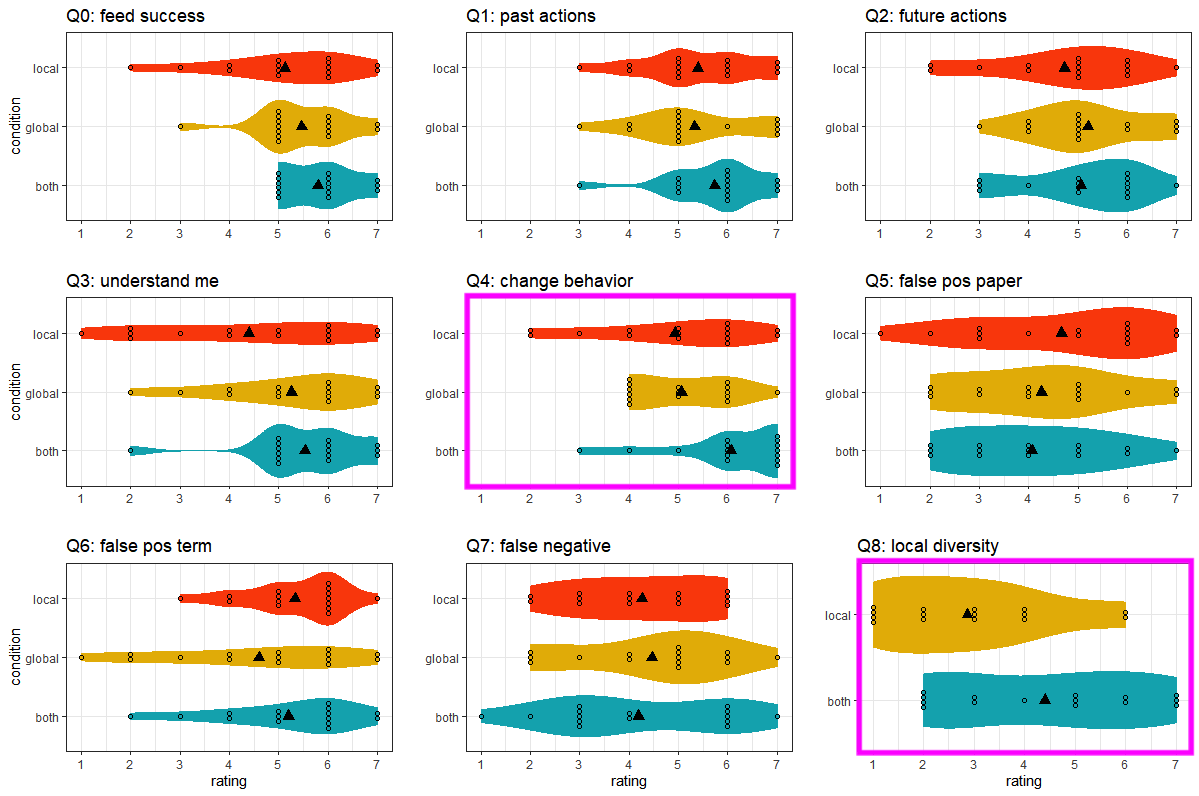}
  \vspace*{-.1in}
  \caption{Study 2 results for each Likert-type question and condition. 1 indicates "strongly disagree," while 7 indicates "strongly agree." Each triangle represents the mean response for the given question and condition, while the circles within each plot represent individual responses. Q4: With both explanations rather than only global (p=0.015, uncorrected, two-tailed Mann-Whitney-Wilcoxon) or only local (p=0.030, uncorrected, two-tailed Mann-Whitney-Wilcoxon), participants were more confident in explaining how they would like the system to behave to be more relevant. Q8: Participants desired less diverse local explanations when the global explanation was also present (p=0.038, uncorrected, two-tailed Mann-Whitney-Wilcoxon).}
  \label{fig:results}
\Description{Violin plots for each condition for each Likert-type question in Study 2. Q0 (feed success): Local has a mean around 5.2, minimum of 2, and maximum of 7. Global has a mean around 5.4, minimum of 3, and maximum of 7. Both has a mean around 5.8, minimum of 5, and maximum of 7. Q1 (past actions): Local has a mean around 5.4, minimum of 3, and maximum of 7. Global has a mean around 5.3, minimum of 3, and maximum of 7. Both has a mean around 5.7, minimum of 3, and a maximum of 7. Q2 (future actions): Local has a mean around 4.7, minimum of 2, and maximum of 7. Global has a mean around 5.2, minimum of 3, and maximum of 7. Both has a mean around 5.1, minimum of 3, and maximum of 7. Q3 (understand me): Local has a mean around 4.4, minimum of 1, and maximum of 7. Global has a mean around 5.3, minimum of 2, and maximum of 7. Both has a mean around 5.6, minimum of 2, and maximum of 7. Q4 (change behavior): Local has a mean around 4.9, minimum of 2, and maximum of 7. Global has a mean around 5.1, minimum of 4, and maximum of 7. Both has a mean around 6.1, minimum of 3, and maximum of 7. Q5 (false pos paper): Local has a mean around 4.7, minimum of 1, and maximum of 7. Global has a mean around 4.3, minimum of 2, and maximum of 7. Both has a mean around 4.1, minimum of 2, and maximum of 7. Q6 (false pos term): Local has a mean around 5.3, minimum of 3, and maximum of 7. Global has a mean around 4.6, minimum of 1, and maximum of 7. Both has a mean around 5.2, minimum of 2, and maximum of 7. Q7 (false negative): Local has a mean around 4.3, minimum of 2, and maximum of 6. Global has a mean around 4.4, minimum of 2, and maximum of 7. Both has a mean around 4.2, minimum of 1, and maximum of 7. Q8 (local diversity): Local has a mean around 2.8, minimum of 1, and maximum of 6. Both has a mean around 4.4, minimum of 2, and maximum of 7.}
\end{figure*}

\subsubsection{Quantitative Results and Discussion}
\label{sec:results}
We organize our discussion of quantitative results around the hypotheses and metrics in Table \ref{tab:metrics}. For the Likert-type questions, we compared the local and global conditions using the within-subjects two-tailed Wilcoxon signed-rank test and the remaining condition pairs using the between-subjects two-tailed Mann-Whitney-Wilcoxon test. Violin plots for these results are presented in Figure \ref{fig:results}. For the log file metrics, we analyzed all condition pairs using a one-way ANOVA test. The significance threshold was p < 0.05. Though all the results were insignificant after Bonferroni corrections, results for \textbf{H4} and \textbf{H7} would be significant without corrections. 

Regarding \textbf{H4}, participants in the local-plus-global condition demonstrated more confidence than the participants in the global (p=0.015, uncorrected) or local (p=0.030, uncorrected) condition in explaining how they would like the system to behave to be more relevant. However, there was no difference indicated between local and global. Thus, the results suggest that \textbf{local and global explanations together are better than either alone for helping users understand how the recommender system can improve}. While similar results have been shown in other machine learning systems \cite{Hohman2019GamutAD,huber2020local}, this is a distinct insight for recommender systems because, unlike those other systems, recommenders do not have objectively correct or incorrect output. The recommender's output is judged and rated according to the user's own standards. This personal form of judgment may benefit more or less from local and global explanations.

In order to create appropriately transparent interactions, a designer needs to know what kinds of information users seek from local explanations. The result for \textbf{H7} suggests that the ideal content of local explanations depends on whether or not a global explanation is present. In particular, \textbf{ participants desired {\em less diverse} and more consistent local explanations when the global explanation was also present} (p=0.038, uncorrected). This may be a consequence of the "explanation-action trade off" \cite{Lee2020ExplanationBasedTO}, which refers to how actionable local explanations in recommender systems must balance two competing goals: 1) returning the most accurate explanations and 2) affording more opportunities for users to adjust the model. The goals are at odds because the most accurate local explanations often share the same terms and thus provide fewer chances for users to adjust the model. We address this in Semantic Sanity by explicitly introducing randomness to diversify the local explanations, as Lee et al. does \cite{Lee2020ExplanationBasedTO}. When local explanations are alone, they are the only means by which users can act on the system, so greater diversity is appreciated by users. In contrast, when an actionable global explanation is also present, users no longer depend on local explanations for adjusting the model and can use them more as a means of explanation, which users may expect to be consistent with the global explanation and thus less random.

\subsubsection{Qualitative Results and Discussion}
In their short-answer responses, \textbf{participants commented more often that they forgot or did not find much use for the local explanations as compared to the global explanation}. Of the 30 participants, 9 mentioned either forgetting local explanations or using them rarely, whereas only one participant mentioned not using the global explanation. However, this difference may be due in part to a user interface design issue, which is described in Section \ref{sec:treatments2}.

Participants also noted that local and global explanations may serve different purposes in terms of research exploration and discovery. Four participants explained that \textbf{the ability to adjust the importance of the global explanation terms was useful to help them avoid unintended bias towards specific authors or topics}. For example, P17 noted, "\emph{The system seemed to be suggesting a particular author and listed that in the feed explanation column. I reduced that so that I could have a more unbiased feed of people I don't often read....}"
Two participants also mentioned that \textbf{the global explanation allowed them to introspect about their own research interests}. For instance, 
P11 commented, "\emph{[Global] gave me a better idea of what my inputs... seemed to have in common.}" 
On the other hand, two participants found \textbf{local explanations were useful for characterizing unexpected interesting papers}. P24 wrote, “\emph{There was a paper suggested to me that I found relevant, but I was also surprised to find it in my recommendation list… [Local] was useful for me to check out why that paper was recommended (so that I can see more such papers!)}.”


\section{Study 3: Controlled User Study}

\subsection{Study Design}
\subsubsection{Hypotheses}
\label{sec:hypotheses2}
Study 2 provided suggestive evidence that both explanations are better than either alone for understanding how the recommender may improve. Study 3's objectives were to confirm this point and to investigate \emph{how} local and global explanations complement one another to help users understand recommender output. Thus, Study 3's hypotheses were as follows. \textbf{H9}: Local is better than global for identifying false positive recommendations, \textbf{H10}: global is better than local for identifying false negative recommendations, and \textbf{H11}: both are better than either alone for understanding how the recommender may improve.
\textbf{H11} exists to confirm the suggestive results from Study 2. \textbf{H9} and \textbf{H10} reflect a framework for how local and global may complement each other to make the recommender more transparent. The hypotheses' associated metrics are provided in Table \ref{tab:study2} and are described further in Section \ref{sec:results2}.

\begin{table*}[ht]
  \caption{Metrics considered in Study 3, with corresponding hypotheses defined in Section \ref{sec:hypotheses2}. The question is a 7-point Likert-type question. The score calculations are described in Section \ref{sec:results2}.}
  \label{tab:study2}
  \begin{tabular}{ p{0.8cm} p{2cm} p{9cm} }
    \toprule
    \textbf{Hypo.} & \textbf{Metric ID} & \textbf{Metric}\\
    \midrule
    H9 & M1 & score on false-positive survey (between -42 and 42)\\
    H10 & M2 & score on false-negative survey (0 or 1)\\
    H11 & Q9 & "I can explain how the system should be updated to be more relevant."\\
  \bottomrule
\end{tabular}
\end{table*}

\subsubsection{Participants and Treatments}
\label{sec:treatments2}
In the same manner as in Study 2, thirty computer-science researchers were recruited and separated into treatments. 
Minimal changes were made to the design of the explanations. Their titles were updated to be purple for emphasis, and they were renamed to better draw users' attention. The local explanations were renamed from "Paper Explanation" to "Why This Paper," and the global explanation was renamed from "Feed Explanation" to "Why This Feed." Also, as is described in Section \ref{sec:proc2}, Study 3's procedure no longer required participants to curate recommendation feeds, so the only clickable buttons were for looking at the explanations and paper abstracts. The remaining buttons 
were still included to provide participants with context for how the recommender system would work overall.

Furthermore, the local-plus-global condition was updated so that local and global explanations were presented separately.
This update was made because, in Study 2, the unified presentation of local and global may have led participants to focus less on local explanations. In the local-plus-global condition, the local explanations could only be opened one-by-one. On the other hand, in the local condition, participants could open as many explanations as they wanted. Perhaps due to this user-interface design, even though participants in the local condition could open all local explanations with a single click, they still opened an individual local explanation 9.3 times on average, while participants in the local-plus-global condition opened an individual local explanation only 2.7 times on average.

When the participants were asked to choose topics of interest for their feeds in Study 2, the feed topics varied largely in breadth and familiarity. This may have hindered our ability to observe significant results in Study 2. As a result, Study 3's participants were randomly assigned to one of two preset feeds for each condition: "misinformation on social media" or "educational technologies for demographically diverse users." These feed topics were chosen based on three criteria. First, in order for participants from varying research areas to engage with the feed, the topic and its explanations needed to use limited jargon. Second, the topic needed to be specific enough that false positives occurred within the top twenty papers of the feed. Third, the topic needed to be broad enough so that a cluster of false negatives emerged. For example, in the "misinformation on social media" feed, true positives were exclusively about \emph{coronavirus-related} misinformation, so any papers discussing misinformation on social media not related to coronavirus formed a cluster of false negatives. The preset feeds were seeded with five papers selected so that the feeds would fit the criteria just mentioned. 

Three annotators classified the top 20 papers of each 250-paper feed as false or true positives and the bottom 50 papers of each feed as false or true negatives, based on the papers’ titles and abstracts. Only papers upon which there was unanimous agreement were added to the pool of papers that the participants could encounter. The original local explanations for each annotated paper were then kept constant, so that no new randomized terms were introduced for diversification.

Subsequently, the twenty-first paper from the "educational technologies for demographically diverse users" feed was added to the pool of papers in order to have enough true positive papers for the study. Also, 
the "misinformation on social media" feed had ten false negatives. Two did not belong to the cluster consisting of papers discussing \emph{non-coronavirus} misinformation on social media. To make sure all participants interacting with this feed would see a false negative from the same cluster, these two false negatives were removed from the pool of papers participants could see. 

\subsubsection{Procedure}
\label{sec:proc2}
Participants first opened a link to the recommender system. For each condition, they then logged into one of two accounts to access a preset feed with six recommendations. Next, we gave them a condition-specific tutorial on using the system. The participants then answered three Google-Forms surveys to address each hypothesis. 

\textbf{H9} was addressed first with a false-positive survey. The survey asked participants to label each of the six paper recommendations in the feed as relevant or not and rate how confident they were in their answers on a 7-point scale. The recommendations were randomly ordered and selected such that half would be false positives. About half of all the true positives had optimal local explanations containing information pertinent to both aspects of the given feed topic. For instance, in the "misinformation on social media" feed, the optimal local explanation may have the term "fake news" related to "misinformation" as well as the term "twitter" related to "social media." To make sure this category of true positive was represented accordingly, one of these true positives was randomly chosen to be included in each participant's feed.

\textbf{H10} was addressed next with a false-negative survey. The survey presented participants with three new paper recommendations for the feed. Two were true positives and one was a false negative. The survey asked participants to rank these papers based on how they believed the recommender system \emph{would rather than should} rank the papers. Ideally, the participant would be able to recognize that the false negative paper would be ranked last by the system. 

Finally, \textbf{H11} was addressed with a survey asking participants to answer the 7-point Likert-type question \textbf{Q9}. The survey also asked participants to explain to a software developer how to make the recommendations more relevant, but we found that participants did not understand this question as intended, so it was discarded.

\subsection{Results and Discussion}
\label{sec:results2}
We organize our discussion of results around the hypotheses and metrics discussed in Table \ref{tab:study2}. The false-positive survey score \textbf{M1} was calculated as follows. For each of the six recommendations, if the participant classified it correctly as relevant or not to the feed topic, 1 multiplied by their confidence (1 to 7) was added to their cumulative score. If the participant classified it incorrectly, -1 multiplied by their confidence was added. The false-negative survey score \textbf{M2} was 1 if the false negative paper was ranked below the two true-positive papers and 0 if not. For \textbf{Q9}, we compared the local and global conditions using the within-subjects two-tailed Wilcoxon signed-rank test and the other condition pairs using the between-subjects two-tailed Mann-Whitney-Wilcoxon test. For \textbf{M1} and \textbf{M2}, we analyzed all condition pairs using a one-way ANOVA test. The significance threshold was p < 0.05. All results were insignificant. 

Regarding \textbf{H11}, the slight change in wording of the Likert-type question \textbf{Q9} in Study 3 as compared to \textbf{Q4} in Study 2 may have implied that, in order to respond affirmatively, the participant needed a more technical rather than merely conceptual understanding of how the system could improve its recommendations. Also, participants may have had more trouble conceptualizing how the system should improve when they did not choose the feed topic. Both of these points would explain the overall lower average response to \textbf{Q9} of 4.83, as compared to the average response to \textbf{Q4} of 5.36.

\begin{figure}[tb]
  \centering
  \includegraphics[width=8cm]{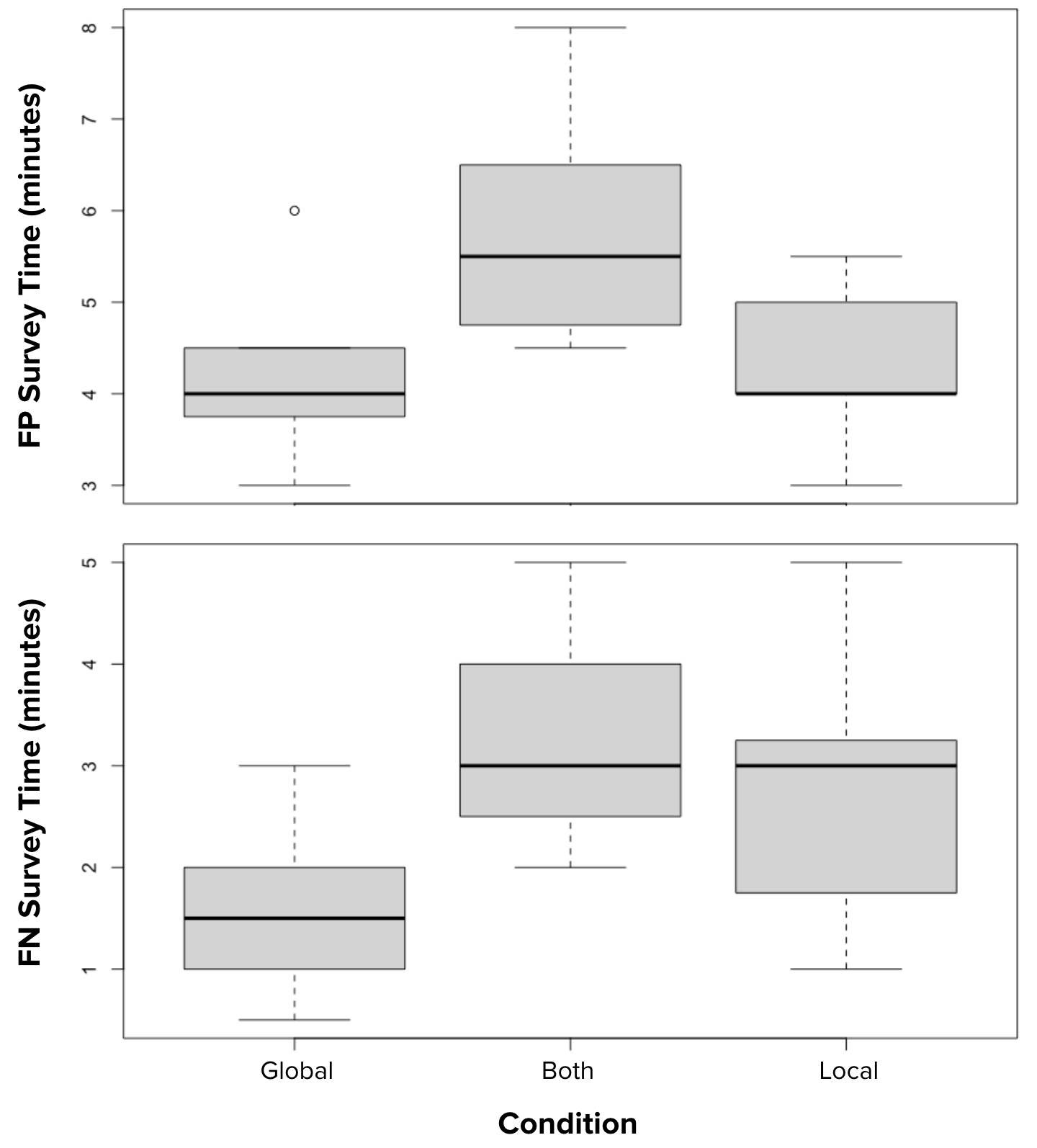}
  \vspace*{-.1in}
  \caption{How much time participants spent on the false-positive (top) and false-negative (bottom) surveys as a function of the explanation condition, under the "misinformation on social media" feed. Top: Participants spent more time on the false-positive survey when both explanations were present as compared to only global (p=0.020, uncorrected, one-way ANOVA) or only local (p=0.045, uncorrected, one-way ANOVA). Bottom: Participants spent less time on the false-negative survey when only global was present as compared to both (p=0.018, uncorrected, one-way ANOVA) or only local (p=0.135, uncorrected, one-way ANOVA). Providing the global explanation alone thus appears more helpful than providing both explanations for identifying false positives and negatives efficiently.}
  \label{fig:boxplots}
  \Description{Top: Boxplots are shown for each of three conditions: global, local-plus-global, and local. The boxplots present how much time in minutes the participants under each condition took to complete the false-positive survey. Global's boxplot has a median around 4 minutes, a first quartile (Q1) around 3.8 minutes, a third quartile (Q3) around 4.5 minutes, a minimum bar around 3 minutes, and a maximum bar around 4.5 minutes. Global's boxplot also has an outlier around 6 minutes. Local-plus-global's boxplot has a median around 5.5 minutes, a Q1 around 4.8 minutes, a Q3 around 6.5 minutes, a minimum bar around 4.5 minutes, and a maximum bar around 8 minutes. Local's boxplot has a median around 4 minutes, a Q1 around 4 minutes, a Q3 around 5 minutes, a minimum bar around 3 minutes, and a maximum bar around 5.5 minutes. Bottom: Boxplots are shown for each of three conditions: global, local-plus-global, and local. The boxplots present how much time in minutes the participants under each condition took to complete the false-negative survey. Global's boxplot has a median around 1.5 minutes, a Q1 around 1 minute, a Q3 around 2 minutes, a minimum bar around 0.5 minutes, and a maximum bar around 3 minutes. Local-plus-global's boxplot has a median around 3 minutes, a Q1 around 2.5 minutes, a Q3 around 4 minutes, a minimum bar around 2 minutes, and a maximum bar around 5 minutes. Local's boxplot has a median around 3 minutes, a Q1 around 1.8 minutes, a Q3 around 3.2 minutes, a minimum bar around 1 minute, and a maximum bar around 5 minutes.}
\end{figure}

Regarding \textbf{H9} and \textbf{H10}, while we did not find any significant differences among the conditions with respect to how well participants identified false positives or negatives, we did observe uncorrected significant differences among the conditions in terms of how \emph{quickly} participants completed the false-positive and false-negative surveys for the "misinformation on social media" feed, as shown in Figure \ref{fig:boxplots}. Each participant's time spent on each survey was rounded to the nearest half-minute. These results are uncorrected because they were not pre-registered for analysis. Twenty-two participants completed a condition using the "misinformation" feed (8 in the global condition, 7 in each of the other conditions). We analyzed all condition pairs for the feed using a one-way ANOVA test followed by a Tukey HSD test. 

With regards to the "misinformation" feed's false-positive survey, participants in the local-plus-global condition completed the survey slower than those in the global (p=0.020, uncorrected) and local (p=0.045, uncorrected) conditions. These results suggest that \textbf{providing both explanations rather than either alone causes users to identify false positives more slowly}. This may simply be due to the fact that there is more information for users to consider when both explanations are available.

With regards to the "misinformation" feed's false-negative survey, participants in the global condition completed it faster than the participants in the local-plus-global condition (p=0.018, uncorrected). Though insignificant, participants in the global condition also completed the survey faster than participants in the local condition (p=0.135, uncorrected). The first result suggests that \textbf{providing only a global explanation rather than both explanations helps users identify false negatives more quickly}. This makes sense for two reasons. Firstly, when both explanations are present, there is more information for users to evaluate. Secondly, in comparison to the local explanations, the global explanation's top terms provide users a straightforward indication of which terms the model may be considering too important or unimportant, which can cause false negatives. 
With only local explanations, users must estimate which terms are most important to the feed by comparing several local explanations.

There are a few possible reasons why we did not observe the same results for the "educational technologies for demographically diverse users" feed. For one, participants generally noted that this topic was more difficult to understand. With respect to the false negative finding, this feed's false negative cluster was the result of an \emph{over-specification} rather than an \emph{irrelevant specification}. The cluster consisted of papers related to educational technologies for \emph{gender-diverse} users. In the global explanation, the only top terms related to diversity were related to ethnic rather than gender diversity. This issue is likely more difficult to note because, unlike the irrelevant specification for "covid" in the "misinformation" feed's global explanation, the term "cultural" in this feed \emph{does} contribute to the feed topic. Terms like "cultural" merely cause the model to overfit to ethnic diversity when terms related to gender diversity should also be included. Thus, global explanations may only help users identify false negatives more quickly when they are the result of an irrelevant specification, as opposed to an over-specification. Lastly, this feed's over-specification in the global explanation was less obvious with fewer related terms than the "misinformation" feed's irrelevant specification.

However, in a follow-up formative study that introduced time constraints for completing the false-positive and false-negative surveys, participants were not evidently better at identifying false positives or negatives in one explanation setting versus another. There are a couple reasons why this could be. For one, computer-science researchers are already accustomed to evaluating the relevance of paper recommendations without explanations, and perhaps often based on titles alone. Explanations may not be useful for identifying false positives and negatives when the recommendations are already sufficiently transparent, especially in a generally lower-stakes situation like browsing research papers. In addition, participants were not necessarily invested in or familiar with the feed topics, as they were not selected by them. As noted in Study 2, there is a trade-off in studying personalized feed topics because they can vary in breadth and familiarity. Nonetheless, given the lack of meaningful results in Study 3 and the inherent individualized nature of recommenders, having participants engage with personal recommendations seems essential to studying recommenders. In a similar vein, since the feed topics were chosen to be accessible to all computer-science researchers, identifying false positives and negatives may have been uncommonly easy.

\section{Conclusion, Limitations, and Future Work}
Following a formative study to determine how content-based local and global explanations should be presented in a research-paper recommender system, we conducted an exploratory study comparing the use of the two explanation approaches in this system. We found evidence suggesting that each explanation type plays a unique role in augmenting the system's transparency and influences how the other is used for understanding the system. Specifically, our results suggest that 
\begin{itemize}
    \item Providing both explanations rather than either alone ensures users reach the best understanding of how the recommender can improve, and 
    \item Users prefer more diverse local explanations when they are presented alone compared to when a global explanation is also available. 
\end{itemize}
The study also provided qualitative evidence that, in the domain of research papers, local and global explanations may be useful for a purpose other than determining recommendation relevance- exploration and discovery of research. 

In a subsequent controlled user study, we investigated \emph{how} local and global explanations may complement one another to help users understand their recommendations, in particular by revealing false positives and false negatives. While we did not find any significant differences between the two explanations in terms of utility in identifying false positives or negatives, we did observe evidence suggesting that 
\begin{itemize}
\item Providing both explanations rather than either alone slows users' identification of false positives, and
\item Providing a global explanation alone rather than both explanations quickens users' identification of false negatives caused by unnecessary specifications.
\end{itemize}
However, a follow-up formative study did not corroborate these findings.

Limitations of this work include that 1) the user studies were small-scale and 2) only one recommendation domain (research papers) and explanation style (content-based) were studied. Future work may study the use of local and global explanations for more opaque recommendations such as author or artist recommendations; an explanation is less necessary when the recommendation itself summarizes its contents, as is the case with paper recommendations. Future research may also explore how these explanations are used in higher-stakes recommendation settings, such as education or healthcare, in which explanations likely bear greater importance. Finally, future work may investigate how local and global explanations are used for purposes other than clarifying recommendation relevance, such as discovery of more diverse recommendations.

\begin{acks}
This research was supported by the University of Washington WRF/Cable Professorship and the Allen Institute for Artificial Intelligence (AI2). The authors thank Matt Latzke and Cecile Nguyen for their advice regarding the project's user-interface design and code respectively. The authors thank all the participants and pilot participants who made this work possible. Finally, the authors thank the anonymous reviewers of this work.
\end{acks}

\bibliographystyle{ACM-Reference-Format}
\bibliography{localglobal}

\end{document}